\documentclass[twocolumn,showpacs,preprintnumbers,amsmath,amssymb]{revtex4-1}%
\usepackage[english]{babel}
\usepackage[T1]{fontenc}
\usepackage{bbm}
\usepackage{epsfig}
\usepackage{psfrag}
\usepackage{color}
\usepackage{amsmath}
\usepackage{amsfonts}
\usepackage{amssymb}
\usepackage{graphicx}%
\providecommand{\U}[1]{\protect\rule{.1in}{.1in}}
\newcommand{\be}{\begin{equation}}
\newcommand{\ee}{\end{equation}}
\newcommand{\bea}{\begin{eqnarray}}
\newcommand{\eea}{\end{eqnarray}}

\begin{document}
\title{The Rubakov-Callan Scattering on the Supergravity Monopole}
\author{Ali~H.~Chamseddine$^{1,2,3}$}
\author{Mikhail~S.~Volkov$^{2}$}
\affiliation{$^{1}$Physics Department, American University of Beirut, LEBANON}
\affiliation{$^{2}$Laboratoire de Math\'{e}matiques et Physique Th\'{e}orique CNRS-UMR
6083, Universit\'{e} de Tours, Parc de Grandmont, 37200 Tours, FRANCE}
\affiliation{$^{3}$LE STUDIUM, Loire Valley Institute for Advanced Studies, Tours and
Orleans, FRANCE}

\begin{abstract}
\vspace{1 cm}

We study small perturbations around the supersymmetric CVMN monopole solution
of the gauged supergravity in D=4. We find that the perturbation spectrum
contains an infinite tower of Coulomb-type bound states both in the bosonic
and fermionic parts of the supergravity multiplet. Due to supersymmetry, the
eigenvalues are the same for the two bosonic parity sectors, as well as for
the fermionic sector. We also find that the fermion scattering on the monopole
is accompanied by  isospin flip. This is analogous to the Rubakov-Callan
effect of monopole catalysis of proton decay and suggests that there could be
a similar effect of catalysis for decay of fermionic systems in supergravity.

\end{abstract}

\maketitle


\ \vspace{1 cm} \

\noindent\textbf{Introduction.--} The Rubakov-Callan effect \cite{RC} can be
viewed as a consequence of the isospin flip for Dirac fermions interacting
with the t'~Hooft-Polyakov magnetic monopole \cite{mon}. When scattered on the
monopole, fermions change their quantum numbers, so that outgoing particles
are not the same as incoming ones. When interacting with systems of bound
fermions, as for example quarks inside a proton, the monopole changes the
quark colors, thus rendering the system unstable. Although not observed in
nature so far, such a monopole catalysis of proton decay is very interesting theoretically.

There are other interesting theoretical effects for magnetic monopoles (see
\cite{R} for a review). For example, scattering of even parity Yang-Mills and
Higgs quanta can resonantly excite the monopole, giving rise to a long living
breathing state \cite{FV}, while the odd parity quanta can be trapped by the
monopole to form bound states \cite{bound}. The monopole can also confine zero
energy fermions, in agreement with the index theorem \cite{JR}.

In this letter we study analogous effects, but in connection to the
supergravity (SUGRA) monopole solution of Chamseddine-Volkov-Maldacena-Nunez
(CVMN). This is an exact solution \cite{CV} of equations of gauged SUGRA in
D=4 that preserves four supersymmetries (SUSY) and contains a Yang-Mills field
whose structure is exactly the same as for the t'~Hooft-Polyakov monopole.
This solution can be promoted to D=10 as a string theory vacuum \cite{CV}, in
which case it can be interpreted as a holographic dual of the N=1
super-Yang-Mills \cite{MN}.

In what follows, we maintain the original interpretation of the solution as
magnetic monopole in D=4 and study its small excitations within the SUGRA
multiplet. We find that boson fluctuations split into two parity sectors and
admit an infinite tower of bound states with the same eigenvalues for both
parities. There are also bound states with exactly the same eigenvalues in the
fermion sector too. We then study the fermion scattering and observe the
isospin flip phenomenon similar to the one discussed by Rubakov and Callan,
even though we do not consider Dirac fermions but interacting spin-3/2 and
spin-1/2 Majorana fields. This suggests that there could be a similar effect
of monopole catalysis of bound fermionic systems in SUGRA.

\noindent\textbf{SUGRA bosons.--} The $N=4$ gauged $SU(2)\times SU(2)$
supergravity of Freedman and Schwarz (FS) \cite{FS} contains the gravitational
field $g_{\mu\nu}$, the axion $\mathbf{a}$, dilaton $\Phi$, and two
non-Abelian gauge fields $A_{\mu}^{a}$ and $B_{\mu}^{a}$ ($a=1,2,3$) with
gauge couplings $e_{A}$ and $e_{B}$, as well as the fermions. One can
consistently truncate the theory to the sector where $B_{\mu}^{a}=e_{B}=0$,
after which one can rescale the remaining fields to achieve the condition
$e_{A}=1$. The action density of the theory then reads
\begin{align}
\mathcal{L} &  =-\frac{1}{4}\,R+\frac{1}{2}\,\partial_{\mu}\Phi\partial^{\mu
}\Phi+\frac{1}{2}\,e^{-4\Phi}\partial_{\mu}\mathbf{a}\partial^{\mu}%
\mathbf{a}\label{0}\\
&  -\frac{1}{4}\,e^{2\Phi}F_{\mu\nu}^{a}F^{a\mu\nu}+\frac{\mathbf{a}}%
{2}\,F_{\mu\nu}^{a}\tilde{F}^{a\mu\nu}+\frac{1}{8}\,e^{-2\Phi}%
+\mathrm{fermions},\nonumber
\end{align}
where $F_{\mu\nu}^{a}=\partial_{\mu}A_{\nu}^{a}-\partial_{\nu}A_{\mu}%
^{a}+\epsilon_{abc}A_{\mu}^{b}A_{\nu}^{c}$ is the gauge field tensor and
$\tilde{F}_{\mu\nu}^{a}$ its dual.
One can consistently set the fermionic fields to zero and study the purely
bosonic fields. Assuming the latter to be spherically symmetric, the most
general line element can be parameterized in spherical coordinates
$t,r,\vartheta,\varphi$ as
\begin{equation}
ds^{2}=2e^{2\Phi}(e^{2\nu}dt^{2}-e^{2\lambda}dr^{2}-U^{2}(d\vartheta^{2}%
+\sin^{2}\vartheta d\varphi^{2})).\label{met}%
\end{equation}
The most general SO(3) invariant gauge field is \cite{FM}
\begin{align}
\mbox{{\bf T}}_{a}A_{\mu}^{a}dx^{\mu} &  =(\Omega_{t}dt+\Omega_{r}%
dr)\mbox{{\bf T}}_{3}+{(\mbox{{\rm k}}_{2}\mbox{{\bf T}}_{1}%
+\mbox{{\rm k}}_{1}\mbox{{\bf T}}_{2})}d\vartheta\nonumber\label{Witten}\\
&  +(\mbox{{\rm k}}_{2}\mbox{{\bf T}}_{2}-\mbox{{\rm k}}_{1}\mbox{{\bf T}}_{1}%
)\sin\vartheta d\varphi+\mbox{{\bf T}}_{3}\cos\vartheta d\varphi
\end{align}
with $[\mbox{{\bf T}}_{a},\mbox{{\bf T}}_{b}]=i\epsilon_{abc}%
\mbox{{\bf T}}_{c}$ being the SU(2) generators. Here $\Phi,\nu,\lambda
,U,\Omega_{t},\Omega_{r},\mbox{{\rm k}}_{1},\mbox{{\rm k}}_{2}$ depend on
$t,r$. In the static case one can set $\nu=\lambda=\mathbf{a}=\Omega
_{t}=\Omega_{r}=\mbox{{\rm k}}_{2}=0$ and $\Phi=\phi(r)$, $\mbox{{\rm k}}_{1}%
=w(r)$, $U={Y}(r)$. The field equations then
admit an exact solution describing the CVMN monopole \cite{CV},
\begin{equation}
w=\frac{r}{\sinh r},~e^{2\phi}=a\,\frac{\sinh r}{{Y}}\,,~{Y}^{2}=2r\coth
r-w^{2}-1,\label{mon}%
\end{equation}
where $a$ as an integration constant. We study time-dependent perturbations of
this solution within the ansatz \eqref{met},\eqref{Witten}, so that to set
\begin{align}
\Phi &  =\phi+\delta\phi,~~~\mbox{{\rm k}}_{1}=w+\delta w,~~~\lambda
=\delta\lambda,~~~\nu=\delta\nu,~~~\nonumber\label{pert}\\
\omega_{1} &  =\delta\omega_{1},~~~\mbox{{\rm k}}_{2}=\delta\mbox{{\rm k}}_{2}%
,~~~\mathbf{a}=\delta\mathbf{a},
\end{align}
where the perturbations depend on $t,r$ and are assumed to be small. The line
element \eqref{met} has a residual symmetry $t\rightarrow t+g(t,r)$,
$r\rightarrow r+f(t,r)$ with $\partial_{r}g=\partial_{t}f$, while the ansatz
\eqref{Witten} is invariant under $\Omega_{a}\rightarrow\Omega_{a}%
+\partial_{a}\alpha(t,r)$, $\mbox{{\rm k}}_{1}+i\mbox{{\rm k}}_{2}\rightarrow
e^{i\alpha}(\mbox{{\rm k}}_{1}+i\mbox{{\rm k}}_{2})$ ($a=t,r$). These
symmetries can be used to impose the gauge conditions $\Omega_{t}=0$ and
$U={Y}(r)$.

Inserting \eqref{pert} into the field equations for the action \eqref{0} and
linearizing with respect to perturbations, the perturbation equations split
into two independent parity groups. In what follows we shall only outline the
results of the complicated detailed calculations which will appear elsewhere.
Let us consider first the even parity group containing $\delta\phi,\delta
w,\delta\nu,\delta\lambda$. The key observation is that the linearized $tr$
component of Einstein equations is a total derivative with respect to time,
which after integrating gives an algebraic relation expressing $\delta\lambda$
in terms of $\delta\phi$ and $\delta w$. The $rr$ component of Einstein
equations then can be resolved with respect to $\delta\nu$, so that the
linearized Yang-Mills and dilaton equations will contain only $\delta\phi$ and
$\delta w$. With $\delta w=e^{i\omega t}e^{-\phi}B_{1}(r)$ and $\delta
\phi=e^{i\omega t}e^{-\phi}B_{2}(r)/Y$, these equations reduce to a
two-channel Schr\"{o}dinger system
\begin{align}
-B_{1}^{\prime\prime}-CB_{1}^{\prime}+V_{11}B_{1}+V_{12}B_{2} &  =\omega
^{2}B_{1},\nonumber\label{s}\\
-B_{2}^{\prime\prime}+CB_{2}^{\prime}+V_{22}B_{2}+V_{21}B_{1} &  =\omega
^{2}B_{2},
\end{align}
where $^{\prime}\equiv\frac{d}{dr}$ and $C,V_{ik}$ are real functions of
$\phi,w,{Y}$ with $V_{21}=V_{12}+C^{\prime}$.

Equations in the odd parity sector contain $\delta\mathbf{a},\delta\Omega_{r}$
and $\delta\mbox{{\rm k}}_{2}$. Setting $\delta\mathbf{a}=\omega\cos(\omega
t)e^{\phi}B_{1}/Y$ and $\delta\Omega_{r}=2\sin(\omega t)e^{-\phi}%
(wYB_{2}+(w^{2}-1)B_{1})/{Y}^{3}$ and also $\delta K_{2}=\sin(\omega
t)e^{-\phi}(e^{-\phi}(we^{\phi}B_{2})^{\prime}/w+2w^{\prime}B_{1}/Y)$, these
equations also assume the Schr\"{o}dinger form \eqref{s}, but the potential
$V_{ik}$ is not the same as in the even-parity case. However, the behavior at
$r\rightarrow\infty$ is similar for both parities, since $C\sim V_{12}\sim
V_{21}\sim e^{-r}\rightarrow0$, so that the system \eqref{s} diagonalizes,
while
\begin{equation}
V_{11}\sim V_{22}\sim\frac{1}{4}-\frac{3}{4r}+O(r^{-2}).
\end{equation}
Since $V_{11}(\infty)=V_{22}(\infty)$, it follows that, unlike for the
t'~Hooft-Polyakov monopole \cite{FV}, the SUGRA monopole does not have
resonant excitations. The spectrum contains scattering states with $\omega
^{2}>1/4$, while for $\omega^{2}<1/4$ there should be an infinite tower of
bound states -- since the potential contains the attractive Coulombian tail.
In the idealized case, if we had exactly $V_{11}=V_{22}=1/4-3/(4r)$ and
$V_{12}=V_{21}=C=0$, the bound states eigenvalues would be $\omega_{n}%
^{2}=1/4-(3/(4n))^{2}$ with $n=1,2,\ldots$ and for every eigenvalue there
would be two different solutions.

Solving Eqs.\eqref{s} numerically, we indeed find bound states with
$\omega^{2}\approx\omega_{n}^{2}$. The double degeneracy present in the
idealized case is lifted for the full system, and for a given $n$ we find two
different solutions with slightly different eigenvalues that we call $n^{+}$
and $n^{-}$, since one of them has $B_{1}\approx B_{2}$ and the other
$B_{1}\approx-B_{2}$. The first ten eigenvalues are shown in Table 1, where it
could be seen that when $n$ increases, they indeed approach the Coulombian
values $\omega_{n}^{2}$. We notice, however, something unusual, since for both
parity values the eigenvalues turn out to be the same, at least up to six
decimal places as shown in the table, even though the potentials $V_{ik}$ are
different. The explanation of this remarkable coincidence is due to
supersymmetry. We next study the fermionic sector.

\begin{table}[ptb]
\begin{tabular}
[c]{|c|c|c|c|}\hline
$n$ & $\frac14-\frac{9}{64n^{2}} $ & even$^{+}$ & even$^{-}$\\\hline
$1$ & $0.10937$ & $0.101710$ & $0.201961$\\
$2$ & $0.21484$ & $0.217134$ & $0.230873$\\
$3$ & $0.23437$ & $0.235150$ & $0.239792$\\
$4$ & $0.24121$ & $0.241552$ & $0.243665$\\
$5$ & $0.24437$ & $0.244553$ & $0.245689$\\\hline
\end{tabular}
\caption{{\protect\small Bound state eigenvalues for Eqs.\eqref{s}.}}%
\end{table}

\noindent\textbf{SUGRA fermions.--} The fermions in the FS model are the
gravitino $\psi_{\mu}$ and gaugino $\chi$. These are Majorana spinors endowed
with an isospin index, so that they have altogether 80 complex components.
Neglecting their self-interactions, their equations of motion read
\cite{FS},\cite{CS}
\begin{align}
\mathcal{R}^{\lambda}  &  \equiv\varepsilon^{\lambda\mu\nu\rho}\gamma
_{5}\gamma_{\mu}\hat{D}_{\nu}\psi_{\rho}-\frac{1}{\sqrt{2}}\,e^{-\phi}%
\sigma^{\lambda\nu}\psi_{\nu}\nonumber\label{ferm}\\
&  +(\frac{i}{{2}}\,e^{\phi}\mathcal{F}-\frac{1}{\sqrt{2}}\,\gamma^{\nu
}\partial_{\nu}\phi-\frac{i}{4}\,e^{-\phi})\gamma^{\lambda}\chi=0,\\
\mathcal{P}  &  \equiv i\gamma^{\mu}{D}_{\mu}\chi+\gamma^{\mu}(\frac{1}%
{\sqrt{2}}\,\gamma^{\rho}\partial_{\rho}\phi+\frac{i}{2}e^{\phi}%
\mathcal{F}-\frac{i}{4}e^{-\phi})\psi_{\mu}=0,\nonumber
\end{align}
where $D_{\mu}=\partial_{\mu}+\frac{1}{4}\,\omega_{~\mu}^{\alpha\beta}%
\gamma_{\alpha}\gamma_{\beta}-\mbox{{\bf T}}^{a}A_{\mu}^{a}$ with
$\omega_{~\mu}^{\alpha\beta}$ being the spin connection and $\hat{D}_{\mu
}=D_{\mu}-\frac{i}{2\sqrt{2}}\,e^{\phi}\mathcal{F}\gamma_{\mu}$ with
$\mathcal{F}={\mbox{{\bf T}}}^{a}F_{\alpha\beta}^{a}\gamma^{\alpha}%
\gamma^{\beta}.$ These equations are invariant under the SUSY transformations
$\psi_{\mu}\rightarrow\psi_{\mu}+\delta_{\epsilon}\psi_{\mu}$, $\chi
\rightarrow\chi+\delta_{\epsilon}\chi$ with
\begin{align}
\delta_{\epsilon}{\psi}_{\mu}  &  =\left(  {D}_{\mu}+\frac{i}{2\sqrt{2}%
}\,e^{\phi}\mathcal{F}\gamma_{\mu}-\frac{i}{4\sqrt{2}}\,e^{-\phi}\gamma_{\mu
}\right)  {\epsilon},~\nonumber\label{SUSY}\\
\delta_{\epsilon}{\chi}  &  =\left(  \frac{i}{\sqrt{2}}\,\gamma^{\mu}%
\partial_{\mu}\phi\,-\frac{1}{2}\,e^{\phi}\mathcal{F}+\frac{1}{4}\,e^{-\phi
}\right)  {\epsilon},
\end{align}
where $\epsilon$ is the spinor SUSY parameter. Due to this invariance, there
exist identity relations between the equations (Bianchi identities)
\begin{align}
D_{\rho}\mathcal{R}^{\rho}  &  +\frac{i}{2\sqrt{2}}\gamma_{\rho}(e^{\phi
}\mathcal{F}-\frac12\,e^{-\phi})\mathcal{R}^{\rho}\nonumber\label{Bianchi}\\
&  -(\frac{i}{\sqrt{2}}\,\gamma^{\rho}\partial_{\rho}\phi+\frac{1}{2}%
\,e^{\phi}\mathcal{F}-\frac{1}{4}\,e^{-\phi})\mathcal{P}{=0}.
\end{align}
It is worth noting that the fermion equations are SUSY invariant iff the boson
background is on-shell. The monopole background \eqref{mon} is on-shell and
moreover it is supersymmetric, since it admits four non-trivial spinors
$\epsilon_{0}$ such that $\delta_{\epsilon_{0}}\psi_{\mu}=\delta_{\epsilon
_{0}}\chi=0$, so that it is invariant under SUSY transformations generated by
$\epsilon_{0}$ \cite{CV}.

Let us collectively denote the bosons by $B$ and fermions by $F$. Their SUSY
variations can be schematically expressed as $\delta_{\epsilon}B=\bar
{\epsilon}F$ and $\delta_{\epsilon}F=D(B)\epsilon$ where $D(B)$ is a covariant
derivative operator. Let us set $F=0$ and $B=B_{0}+\delta B$ where $B_{0}$ is
the monopole and $\delta B$ its perturbation. Since this configuration is
on-shell, so will be its SUSY variations. Let us consider the variations
induced by the Killing spinors $\epsilon_{0}$. One has $\delta_{\epsilon_{0}%
}B=0$, while
\begin{equation}
\delta_{\epsilon_{0}}F=D(B_{0}+\delta B)\epsilon_{0}\approx D(B_{0}%
)\epsilon_{0}+D(\delta B)\epsilon_{0}=D(\delta B)\epsilon_{0},\label{del}%
\end{equation}
where we used the fact the the background is SUSY-invariant i.e.
$D(B_{0})\epsilon_{0}=0$. By construction, \eqref{del} should fulfill the
fermion equations. Therefore, to every perturbative solution $\delta B$ in the
bosonic sector there corresponds a solution $D(\delta B)\epsilon_{0}$ in the
fermionic sector. More precisely, it is given by \eqref{SUSY} with
$\epsilon=\epsilon_{0}$.

It then follows that the fermionic equations \eqref{ferm} should contain all
solutions of the bosonic equations. In particular, they should have the same
bound state spectrum. To verify this, we should solve the system of $80$
spinor equations \eqref{ferm}. As a first step, we impose the symmetry
condition on spinors, whose total angular momentum $J=L+S+I$ consists of the
orbital part $L$, spin $S$ and isospin $I$. Since $I=1/2$, both for $S=3/2$
(gravitino) and $S=1/2$ (gaugino), there are integer values of $L$ giving
$J=0$. Therefore, both the gravitino and gaugino could form spherically
symmetric states. When we restrict to $J=0$, the dependence on the angles
$\vartheta,\varphi$ separates, and the fermionic system \eqref{ferm} reduces
to 32 equations for 32 complex functions of $t$ and $r$. In addition, the
Majorana condition eliminates half of the degrees of freedom, so that we are
left with only 16 equations for 16 complex amplitudes.

As a consistency check, we verify that these equations admit \eqref{SUSY} as
solutions (provided that $\epsilon$ also has $J=0$) and that they fulfill the
Bianchi identities \eqref{Bianchi}. Next, we verify that if $\delta B$ is a
solution of the Schr\"{o}dinger problem \eqref{s}, either for even or for odd
parity, then $D_{\mu}(\delta B)\epsilon_{0}$ fulfills the fermion equations.
After this we are confident that our equations are correct, and so we proceed
to solve them. In order to fix the gauge, we impose the condition
\begin{equation}
\gamma^{0}\psi_{0}+\frac{i}{\sqrt{2}}\,\chi=0,
\end{equation}
{which removes all time-dependent pure gauge modes} and eliminates 4 complex
amplitudes out of 16 yielding {16} equations for {12} functions. It turns out
that 4 of the 16 equations are algebraic and can be used to express 4
functions in terms of the other 8. As a result, there remain {12} equations
for {8} functions. Assuming the harmonic time dependence $e^{i\omega t}$ for
the spinors, the time variable separates. Taking linear combinations, we find
that only 8 of the remaining 12 equations are differential while the remaining
4 are algebraic constraints.
We check then that differentiating these constraints and using the 8
differential equations to eliminate the derivatives do not lead to new
constraints. We can therefore resolve the constraints to express four
amplitudes in terms of the other four, so that everything reduces to just four
first order differential equations with real coefficients. Converting them to
two second order equations, we finally obtain
\begin{align}
-F_{1}^{\prime\prime}+(U_{11}-{\omega^{2}})F_{1}+{\omega}U_{12}F_{2} &
=0,\nonumber\label{ss}\\
-F_{2}^{\prime\prime}+(U_{22}-{\omega^{2}})F_{2}+{\omega}U_{21}F_{1} &  =0,
\end{align}
where $U_{ik}$ are real functions of the background amplitudes $w,\phi,{Y}$.
Summarizing, we managed to reduce the 80 fermion equations \eqref{ferm} to two
second order equations \eqref{ss}. These equations are solved to determine
$F_{1}(r),F_{2}(r)$, in terms of which all components of $\psi_{\mu},\chi$ can
then be expressed.

Before proceeding, we analyze the asymptotic behavior of solutions and find
that for $r\rightarrow\infty$ one has $F_{1}\sim F_{2}\sim e^{-3\phi/2-kr}$
with $k=\sqrt{1-\omega^{2}/4}$. Since $\phi\sim r/2$ at large $r$, it follows
that the solutions are always exponentially suppressed at infinity,
irrespective of the value of $\omega$. Fermions are therefore always localized
around the monopole and cannot escape to infinity, as if they had no
scattering states. However, this seems to be a purely kinematical effect,
since passing to the string frame $ds^{2}\rightarrow e^{-2\phi}ds^{2}$ changes
the spinors as $F\rightarrow e^{3\phi/2}F$ so that they oscillate at infinity
for $\omega^{2}>1/4$ and are exponentially suppressed for $\omega^{2}<1/4$. We
then solve equations \eqref{ss} numerically looking for bound states with
$\omega^{2}<1/4$ and obtain {exactly the same} eigenvalues as those given in
Tab.1.
This result is of course natural, since we know that the fermion equations
contain all solutions of the bosonic sector due to the map $\delta
B\rightarrow D(\delta B)\epsilon_{0}$.

In the bosonic sector there are two different sets of equations, one for even
parity and one for odd parity. The eigenvalues are the same in both cases, so
that every eigenvalue is doubly degenerate. In the fermionic sector we obtain
only one set of equations, and for every eigenvalue we find only one solution.
This solution should therefore be the image of both even parity and odd parity
bosonic modes. When explicitly calculating $F_{1},F_{2}$ corresponding to
$D(\delta B)\epsilon_{0}$, we obtain the same result regardless of whether
$\delta B$ has even or odd parity,
\begin{equation}
B_{1}^{\mathrm{even}},B_{2}^{\mathrm{even}}\rightarrow F_{1},F_{2}\leftarrow
B_{1}^{\mathrm{odd}},B_{2}^{\mathrm{odd}}.
\end{equation}
Therefore, the even parity and odd parity bosonic sectors can be related to
each other by a change of variables via the fermion sector. This finally
explains why the spectrum is the same for both parities.

\noindent\textbf{Fermion scattering.--} We are now ready to analyze the
scattering problem. Let us consider a wave ingoing from infinity in the
$F_{1}$ channel of system \eqref{ss}. It will approach the monopole core,
where it will excite the amplitude also in the $F_{2}$ channel. The wave
reflected from the center will go back to infinity being distributed between
both channels, so that for $r\rightarrow\infty$ one will have
\begin{equation}
F_{1}(r)\rightarrow e^{ikr}+a_{1}e^{-ikr},~~~~~~F_{2}(r)\rightarrow
a_{2}e^{-ikr}, \label{refl}%
\end{equation}
where $a_{1},a_{2}$ are complex functions of ${\omega}$. \begin{figure}[th]
\hbox to \linewidth{ \hss
	
	\psfrag{x}{$\omega$}
	\psfrag{a}{$\frac{|a_2|}{\sqrt{|a_1|^2+|a_2|^2}}$}
	\psfrag{b}{$\frac{|a_1|}{\sqrt{|a_1|^2+|a_2|^2}}$}
	\resizebox{8cm}{5cm}{\includegraphics{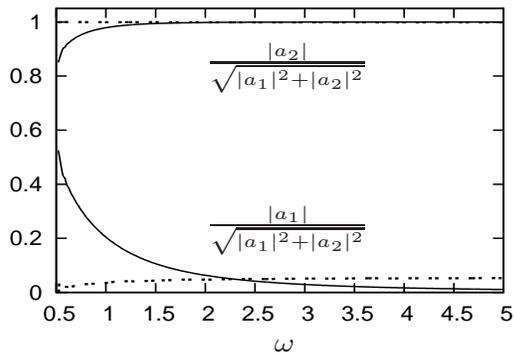}}
\hspace{1mm}
	
\hspace{1mm}
\hss}
\caption{{\protect\small The reflection and transmission coefficients
$a_{1},a_{2}$ against $\omega$ if the incident wave is in the $F_{1}$ channel
(solid lines), or in the $F_{2}$ channel (dotted lines).}}%
\end{figure} Solving equations \eqref{ss} with such boundary conditions shows
that the reflection coefficient $a_{1}({\omega})$ rapidly approaches zero (see
Fig.1). Therefore, the $F_{1}$ ingoing wave gets converted into the $F_{2}$
outgoing wave when scattered on the monopole. We call this phenomenon isospin
flip, since it is very similar to the isospin flip for the Dirac fermions
scattered on the t'~Hooft-Polyakov monopole. The only difference is that in
the latter case the fermions do not encounter a centrifugal barrier \cite{RC},
whereas our equations \eqref{ss} turn out to contain the $2/r^{2}$ centrifugal
term at small $r$. As a result, we do not always have a 100$\%$ isospin flip,
but a flip rapidly approaching 100$\%$ as the energy increases.

If the wave incident from infinity is in the $F_{2}$ channel, then the
asymptotic behavior for $r\rightarrow\infty$ is given by \eqref{refl} with
$F_{1}$ and $F_{2}$ interchanged. As seen in Fig.1, in this case $a_{1}%
(\omega)$ does not tend to zero when $\omega$ increases, but it is always very
small and approaches $\approx0.05$, so that the isospin flip is $\approx95\%$.

A consequence of the isospin flip on the t'~Hooft-Polyakov monopole is the
Rubakov-Callan effect of monopole catalysis of proton decay \cite{RC}. It is
therefore suggestive that the CVMN monopole could similarly catalyze the decay
of fermionic systems in SUGRA.

One can also consider the fermion zero modes. For example, differentiating the
background \eqref{mon} with respect to the scale parameter $a$ gives a boson
zero mode $\delta B$, which can be converted to the fermion mode via $\delta
B\rightarrow D(\delta B)\epsilon_{0}$. However, since our spinors are Majorana
and not Weyl, the relation between fermion zero modes and the monopole
topology is less clear than in the standard case \cite{JR}. These and other
issues will be explored in a detailed publication.

\noindent\textbf{Acknowledgments.--} We thank Y.~Shnir for discussions. The
research of A. H. C. is supported in part by the National Science Foundation
under Grant No. Phys-0854779.

\end{document}